\documentclass[1p]{elsarticle}
\usepackage{amsmath}
\usepackage{amsfonts}
\usepackage{graphicx}
\usepackage{caption}

\date{\today}

\def\bA{\mathbf A}

\def\bE{\mathbf E}
\def\bH{\mathbf H}
\def\bJ{\mathbf J}

\def\bS{\mathbf S}

\def\bx{\mathbf x}

\def\bd{\mathbf d}

\def\bk{\mathbf k}

\def\b0{\mathbf 0}
\def\bTheta{\mathbf \Theta}

\begin{document}






\begin{frontmatter}

\title{Superluminal, subluminal, and negative velocities in free-space 
electromagnetic propagation}
\author{Neil V. Budko}
\address{Laboratory of Electromagnetic Research, Faculty of Electrical Engineering, Mathematics and Computer Science,
Delft University of Technology, Mekelweg 4, 2628 CD Delft, The Netherlands}
\ead{n.v.budko@tudelft.nl}

\begin{abstract}
In this Chapter the time-domain analysis of the velocity of the electromagnetic
field pulses generated by a spatially compact source in free space is presented. 
Recent simulations
and measurements of anomalous superluminal, subluminal, and negative velocities are
discussed. It is shown that such velocities are local and instantaneous in nature and
do not violate either causality or special relativity. Although these effects are mainly
confined to the near- and intermediate-field zones, some of them seem paradoxical and 
still lack adequate physical interpretation.
\end{abstract} 

\end{frontmatter}

\section{Introduction}
Historically, the fact that the electromagnetic field propagates in vacuum with the speed of light
$c=299792458$~m/s was used to establish the electromagnetic (wave) nature of light and the 
correctness of Maxwell's equations. The invariance of the speed of light with respect
to inertial reference frames has lead to the special relativity theory and its idea of 
space-time and was used to fix the standard of length once and for all, Giacomo~(1984). 
The realization that nothing can travel faster than light was as profound as the 
deduction of the impossibility of perpetual motion from the second 
law of thermodynamics. Often, however, measurements or simulations are reported
which seem to demonstrate velocities greater or smaller than what is expected
from light. 

It has been known for a long time that the notion of velocity becomes dubious when a wave propagates 
in a medium with strong anomalous dispersion, see Landau~\&~Lifshitz~(1963). 
The envelope of a wave in such a medium may 
deform during the propagation so that it may seem to travel faster than light, 
slow down, completely stop, or even travel with a negative velocity (towards the source). 
Measurements confirming these strange effects were
reported by Segard and Macke (1985), Steinberg {\it et al} (1993), Hau {\it et al} (1998), Wang,~L.~J. {\it et al} (2000), Bajcsy {\it et al} (2003), 
and Wang, H. {\it et al} (2007). 

With recent advances in composites, materials have been designed 
(notably, photonic crystals and metamaterials) that mimic
microscopic dispersion on a macroscopic level in an effective sense,
i.e., their effective constitutive parameters exhibit anomalous dispersion in
a prescibed frequency band. There too unusual velocities were predicted and
measured, see e.g. Spielman {\it et al} (1994) and Woodley {\it et al} (2004).

Most surprisingly, however, strange velocities have been predicted and observed with free-space 
waves as well.
For example, Bialynicki-Birula {\it et al} (2002) and Cirone {\it et al} (2002) discuss the 
evolution of the wavefunction of 
a free quantum particle with zero angular momentum. They observe that the initial propagation 
of the average radial position for a ring-shaped wave 
packet may be negative. This effect is specific to the two-dimensional case, 
and is related to the violation of the Huygens principle in the space of even dimensions.

Recently free-space measurements of anomalous velocities with the
electromagnetic field were reported by Ranfagni {\it et al} (1993, 1996) and
Mugnai {\it et al} (2000). The authors have detected superluminal propagation
in controlled electromagnetic beams with specific transverse profiles. 
Understandably these reports steered up some discussion and controversy,
see e.g. Heitmann~\&~Nimtz (1994), Diener (1996), Porras (1999), Wynne (2002),
and Zamboni-Rached~\&~Recami (2008).
These anomalous effects, especially in free-space propagation, indicate 
an intrinsic ambiguity of the notions of velocity and speed when 
applied to the complex phenomenon of electromagnetic radiation. On the one hand, we 
have a well-defined constant of nature $c=299792458$~m/s related to the electric
and magnetic properties of vacuum as $c=1/\sqrt{\varepsilon_{0}\mu_{0}}$. On the
other hand, these constants show up as mere parameters in the Maxwell equations
and do not directly refer to any kinematic process, i.e., something like a flight 
of a projectile. In fact, any kinematic analogy at all is extremely far-fetched 
with respect to monochromatic plane waves,
where most of the wave-speed/velocity theory has been developed so far, and the 
notions of phase and group velocities were introduced, see e.g. 
Landau~\&~Lifshitz (1963) for a discussion of linear dispersive media. 
Considerably less was done in the time domain, i.e. directly 
with pulsed signals where the kinematic interpretation would seem to be more natural. 
In this chapter we shall exploit such a time-domain approach. 

The operational notion of velocity requires a displacement of something in space 
during a certain interval of time. The ratio of the spatial displacement to the 
corresponding time interval gives the component of the velocity along a particular
spatial direction, provided that this ratio does not depend of the measurement
interval. This is only true for the motion with a constant-velocity. Otherwise 
the limit of a vanishingly small interval must be taken, thus defining an instantaneous velocity.       

In a homogeneous medium without dispersion the light is believed to travel with a constant velocity.
Moreover, if the source of light is localized, then we expect the wave to 
propagate strictly outwards. Hence, to measure the speed of light it is 
sufficient to place two receivers in line with the source, one further away than
the other, and mark the time instants when the light passes each of them.
Dividing the distance between the receivers by the difference between the
detection times we should get the speed of light within the usual measurement errors.      

In the next section an experimental implementation of this two-point measurement with
a short microwave pulse radiated by a small antenna in free space is considered.
It gives a remarkable result, strikingly similar to the one reported by Ranfagni {\it et al} (1993, 1996) and
Mugnai {\it et al} (2000). Namely, close to the source the electromagnetic pulse
appears to travel faster than light, whereas further away the velocity of outward 
propagation approaches the speed of light. Much of the remainder of this chapter is devoted to 
a theoretical explanation of this phenomenon from the time-domain point of view. 
We shall see that the dynamics of
an electromagnetic pulse is not a simple outward propagation even in free space.
A careful examination of the electromagnetic causality principle shows that this principle applies 
only to the relation between the source -- current density -- and the field, and to 
nothing else. Moreover, this relation is complex enough to allow not only
faster than light pulses, but also pulses with apparently negative velocities, i.e.,
traveling towards the source.
The latter effect was recently experimentally confirmed by the author, Budko (2009). 

Thus, in general, the electromagnetic 
field does not propagate at the speed of light. In fact, 
an instantaneous local velocity seems to be an all together more appropriate concept.
For a pulse this local velocity approaches the speed of
light from above as one moves further from the source. However, the wavefront, 
i.e., a hard to measure boundary 
between the field and the region of space which it has not reached yet, always travels
at the speed of light. We shall also be looking into the power flow from a small
transient source. Although, the time-integrated power flow is always positive, i.e.,
away from the source, in the near-field zone the instantaneous power may flow
towards the source for some time. Finally, we consider the propagation velocity 
of the electromagnetic energy, defined as the ratio of the local instantaneous 
power flow and instantaneous energy density. This velocity never exceeds the 
speed of light in magnitude,
however, may also become subluminal and even negative.

\section{One failed demonstration}
 Outside the source, in vacuum or air, the electric field 
strength $\bE(\bx,t)$ satisfies the homogeneous wave equation  
\begin{align}
\label{eq:WaveHomogeneous}
\Delta \bE(\bx,t)-\frac{1}{c^2}\partial_{t}^{2}\bE(\bx,t)=\b0.
\end{align}
This is a {\it scalar} wave equation satisfied by each Cartesian component of $\bE$ individually.
In the one-dimensional case 
\begin{align}
\label{eq:OneDWave}
\partial_{x}^{2}u(x,t)-\frac{1}{c^{2}}\partial_{t}^{2}u(x,t)=0,
\end{align}
with the initial conditions
\begin{align}
\label{eq:OneDInit}
u(x,0)=f(x),\;\;\;\partial_{t}u(x,0)=g(x),
\end{align}
the wave equation has the general (d'Alembert) solution of the form
\begin{align}
\label{eq:Dalamber}
u(x,t)=\frac{f(x-ct)+f(x+ct)}{2}+\frac{1}{2c}\int_{x-ct}^{x+ct}g(x')\,{\rm d}x'.
\end{align}
Here one can see that for every $t>0$ the first term represents two copies of the initial 
field distribution $f(x)$ reduced by half in magnitude and shifted in space by 
the amount $\pm ct$. This shift is proportional to $c$, which thus has not only 
the dimensions, but also the obvious interpretation as the speed of propagation.

Although things are more complicated in the three-dimensional case, the interpretation 
of coefficient $c$ as the speed of propagation follows, in particular, from the spherical-wave 
solution of Eq.~(\ref{eq:WaveHomogeneous}) with a point-source. Hence, to demonstrate
the finite and well-defined speed of electromagnetic waves it is sufficient to 
generate a current pulse in a relatively small antenna and show that the electromagnetic 
field spreads out as an approximately spherical wave with positive radial 
velocity. We use here the usual distinction between the velocity and the speed,
where the latter is the always positive magnitude of the former. Throughout the text we mainly discuss
a particular Cartesian component of the velocity vector -- along the outwards direction with respect 
to the source. In the first set of experiments this component is ``chosen'' by the measurement 
procedure itself. Further, when we turn our attention to the power flow, it appears to be the 
only component present in our measurement configuration (for the chosen mutual orientation 
of antennas). As we are talking about a Cartesian component of the velocity, 
its value can be positive, negative or zero.

\begin{figure}[t]
        \begin{center}
	\includegraphics[width=0.7\textwidth]{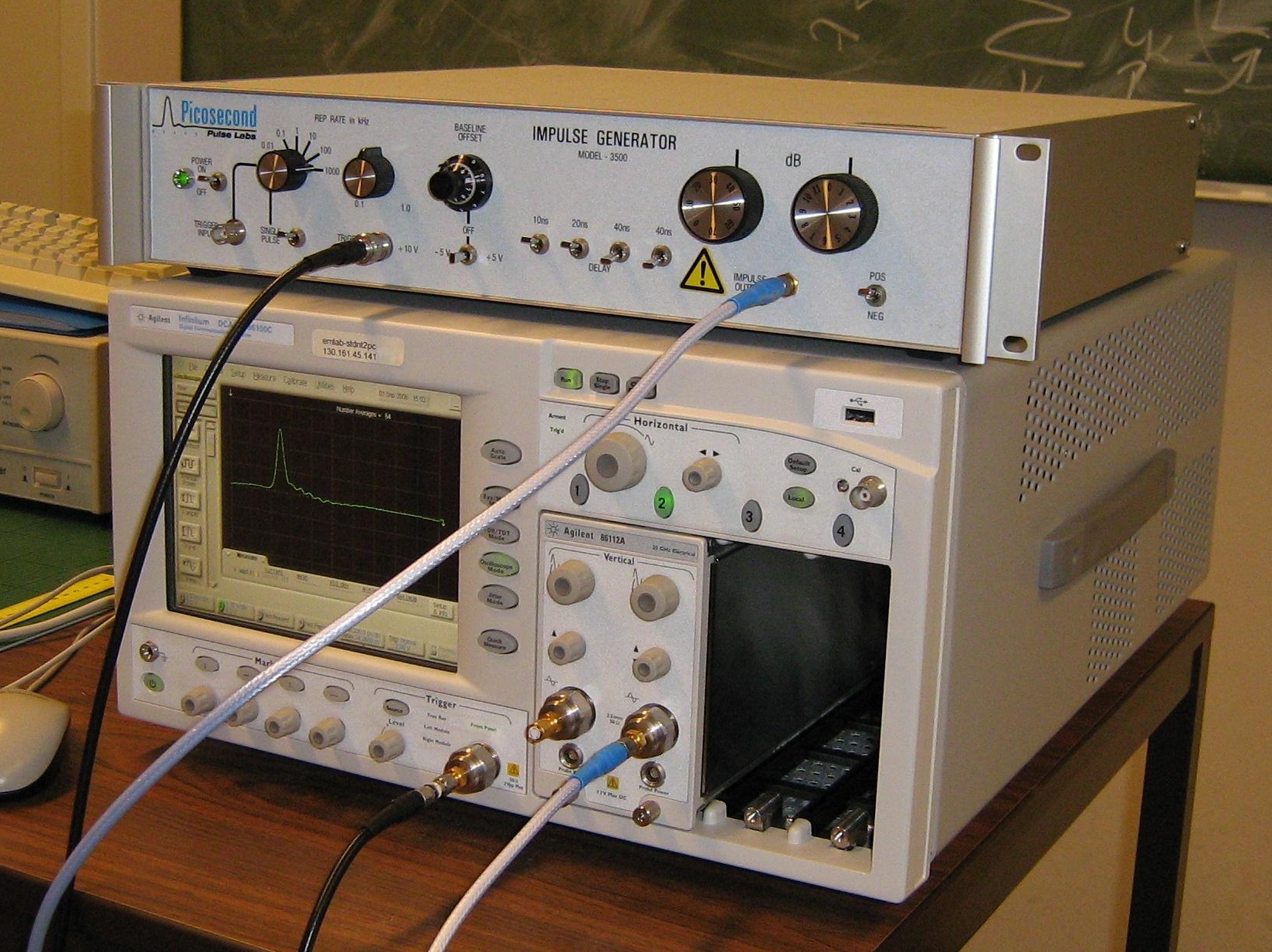}
	\includegraphics[width=0.7\textwidth]{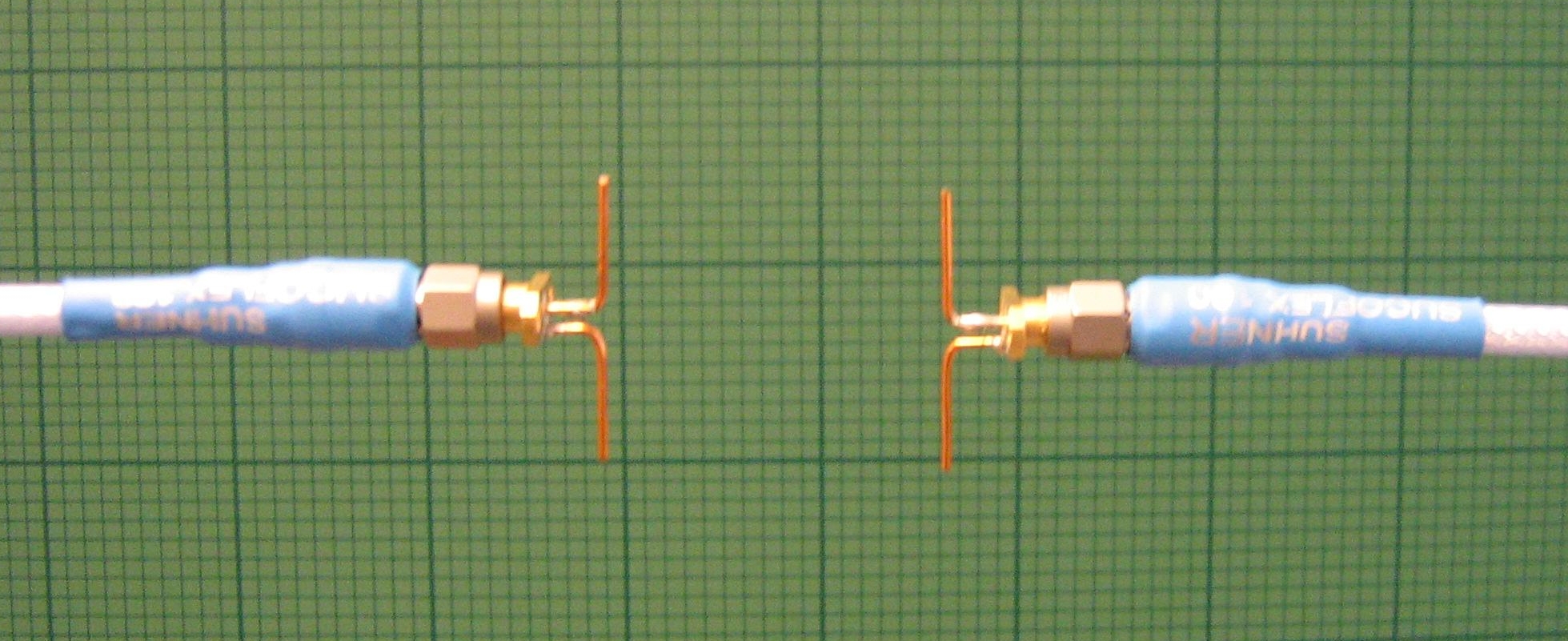}
        \end{center}
	\captionof{figure}{Equipment used to measure the velocity of electromagnetic waves.
Upper photo: impulse generator on top of a sampling oscilloscope. Lower photo:  
dipole antennas attached to broadband co-axial cables (dipoles are less than 3~cm long).}
	\label{fig:ExperimentalSetup}
\end{figure}

Modern microwave equipment allows to carry out direct time-domain two-point measurements in the classroom. 
For instance, the experimental setup shown 
in Fig.~\ref{fig:ExperimentalSetup} was used in the introductory lecture of the graduate
course on Electromagnetics at the Delft University of Technology. 
The author's plan was to follow the following simple routine: 
\begin{enumerate}
\item{receive the same signal at two locations, one further away from the source than
the other;}
\item{choose some characteristic feature of the received waveform, say, the first maximum;} 
\item{measure the absolute arrival times of this feature at two locations;}
\item{divide the difference between the distances of the two receivers 
to the source by the difference between the corresponding arrival times.}
\end{enumerate}
The result of the last calculation must be equal to the speed of light in vacuum.
The required picosecond 
temporal resolution is routinely achieved nowadays with an impulse generator 
producing almost identical impulses and trigger signals at a
precisely controlled repetition rate. The remaining jitter is taken care of by averaging
over many realization of the pulse.
The absolute arrival times here are measured in the same (stationary) reference frame, and in fact
using the same well-calibrated device -- the sampling oscilloscope -- with time-zero set 
at a fixed time-interval with respect to the trigger.  

Let the relative (radial) distance between the receiver locations be 12~cm. 
Then, the relative difference in arrival times must be
\begin{align}
\label{eq:12cm}
\Delta t=\frac{0.12\;{\rm m}}{299792458\; {\rm m/s}}=0.40028\;{\rm ns}.
\end{align}

\begin{figure}[t]
	\includegraphics[width=1.00\textwidth]{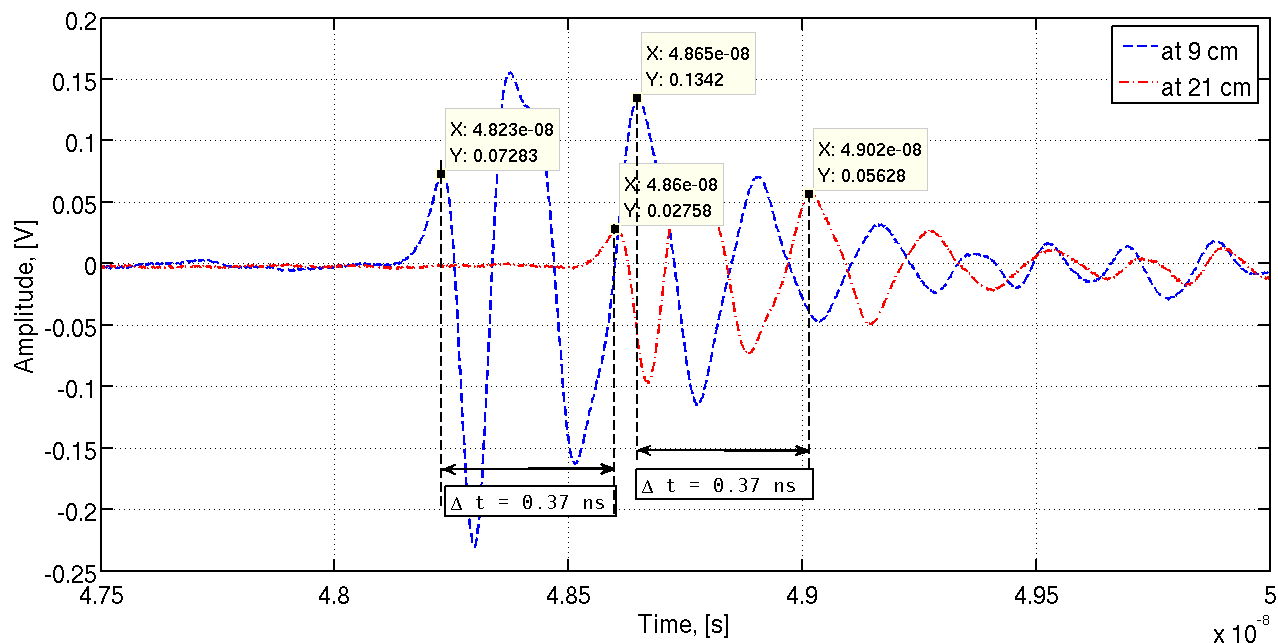}
	\captionof{figure}{Superluminal propagation as seen in two-point near-field measurements. 
                           These signals were measured at 9 and 21 cm from the source 
                           (12 cm spatial separation). The arrival-time difference is 
                           too small: $\Delta t=0.37$~ns.}
	\label{fig:fastnear}
\end{figure}

\begin{figure}[t]
	\includegraphics[width=1.00\textwidth]{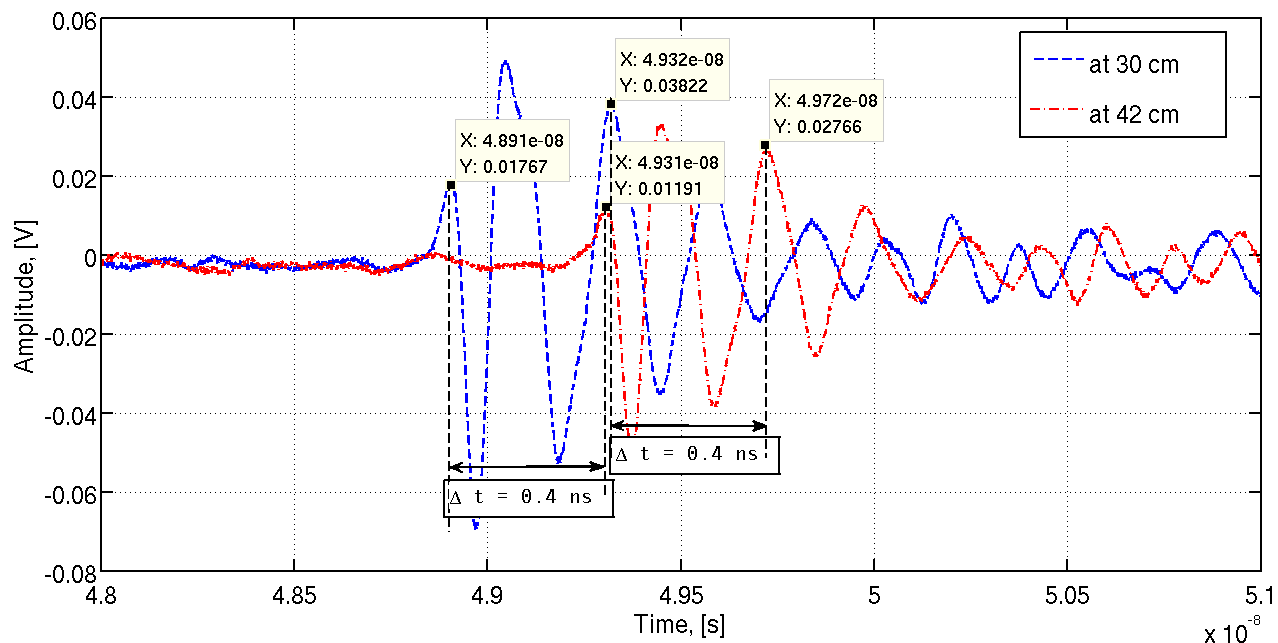}
	\captionof{figure}{Further away from the source propagation happens at the speed of light. 
                           These signals were measured at 30 and 42 cm from the source 
                           (same 12 cm spatial separation as in Fig.~\ref{fig:fastnear}). The arrival-time difference is 
                           now as expected: $\Delta t=0.4$~ns.}
	\label{fig:normalfar}
\end{figure}

In Fig.~\ref{fig:fastnear} and Fig.~\ref{fig:normalfar} we see the results of actual measurements.
The first measurement is performed near the source: receivers at 9~cm and 21~cm. In the second measurement
the receivers are at 30~cm and 42~cm. Thus, both figures correspond to the 12~cm relative spatial separation of the receivers. 
As we have some freedom in choosing a waveform feature to mark the arrival times, we compare the results obtained 
using the first and the third maximum of the waveform.
The data tips shown in the figures contain the absolute time values in seconds ($X$-axis data) at the extrema together 
with the corresponding voltage values (irrelevant to us here). The absolute time values are used to calculate $\Delta t$.

To the surprise and confusion of students only the second set of measurements, 
for the receivers at 30~cm and 42~cm, gives the correct answer, i.e., $\Delta t=0.4$~ns. The 
relative difference in the arrival times closer to the source is consistently smaller, e.g. $\Delta t=0.37$~ns in Fig.~\ref{fig:fastnear},
showing {\it superluminal} propagation! 

Nothing can travel faster than light. Hence, either the measurements are completely wrong
or the maxima of a waveform cannot be regarded as true physical entities, i.e., cannot be
the carriers of information. On the other hand, such waveform features are routinely
used to transmit data and deduce distances to objects (travel time tomography), see e.g. 
Valle {\it et al} (1999) and Skolnik (2001).  
Could it be that the two-point measurement procedure does not determine the velocity?
Yet, this procedure represents the very definition of velocity.
It turns out that the resolution of this puzzle can be found in the classical radiation
formula relating the current in the source and the field at the receiver.

\section{Causality, compatibility, continuity, and gauge fixing.}
The Maxwell equations in vacuum
\begin{align}
\label{eq:Maxwell}
\begin{split}
-\nabla\times\bH(\bx,t)+\varepsilon_{0}\partial_{t}\bE(\bx,t)&=-\bJ(\bx,t),
\\
\nabla\times\bE(\bx,t)+\mu_{0}\partial_{t}\bH(\bx,t)&=\b0,
\end{split}
\end{align}
give the {\it local} relation between the source, current density $\bJ$, 
and the fields, electric and magnetic field strengths $\bE$ and $\bH$.
To relate the source at one location to the fields at another
one has to solve the Maxwell equations, i.e., derive
the {\it radiation formula}. Explicit solutions of this kind can be obtained only 
in some elementary cases, such as the homogeneous background medium we consider
here. We are particularly interested in the way the speed
of light $c=1/\sqrt{\varepsilon_{0}\mu_{0}}$ enters the radiation formula, how exactly
it reflects the causality of the radiation process, and whether this causality
extends to such obvious and practically important features of the waveform as its extrema.

Causality is an assumption that the fields $\bE$ and $\bH$ are 
caused by the current $\bJ$.
We must ensure therefore that the left- and right-hand sides of the Maxwell 
equations are {\it compatible}, i.e., all sources are accounted for by the current 
density $\bJ$.   
To see what it really means let us reduce the Maxwell's system of
two first-order equations to the second-order vector wave equation:
\begin{align}
\label{eq:VectorWave}
\nabla\times\nabla\times\bE(\bx,t)+\frac{1}{c^{2}}\partial_{t}^{2}\bE(\bx,t)=
-\mu_{0}\partial_{t}\bJ(\bx,t).
\end{align} 
From the first of the Maxwell equations we derive the following compatibility relation:
\begin{align}
\label{eq:Compatibility}
\varepsilon_{0}\partial_{t}\nabla\cdot\bE(\bx,t)=-\nabla\cdot\bJ(\bx,t).
\end{align}
Temporal causality is often introduced as an assumption that no field existed before the 
initial switch-on moment $t_{0}$. Hence, we can re-write the local (in time)  
relation (\ref{eq:Compatibility}) as a time-integrated formula
\begin{align}
\label{eq:IntCompatibility}
\nabla\cdot\bE(\bx,t)=-\frac{1}{\varepsilon_{0}}\nabla\cdot\int_{t_{0}}^{t}\bJ(\bx,t')\,{\rm d}t'.
\end{align}
For an ``adiabatic'' source one should take $t_{0}\rightarrow -\infty$.
Now, using this expression and identity $\nabla\times\nabla\times\bE=\nabla\nabla\cdot\bE-\Delta\bE$ 
we reduce the vector wave equation (\ref{eq:VectorWave}) to the
usual scalar wave equation
\begin{align}
\label{eq:ScalarWaveGradDiv}
\Delta\bE(\bx,t)-\frac{1}{c^{2}}\partial_{t}^{2}\bE(\bx,t)=
-\frac{1}{\varepsilon_{0}}\nabla\nabla\cdot\int_{t_{0}}^{t}\bJ(\bx,t')\,{\rm d}t'
+\mu_{0}\partial_{t}\bJ(\bx,t),
\end{align}
where the left-hand side is in its standard form (\ref{eq:WaveHomogeneous}) and the
source term is explicitly identified. 

The compatibility relation (\ref{eq:IntCompatibility}),
is not only a key step in transforming the Maxwell equations into 
the wave equation (\ref{eq:ScalarWaveGradDiv}), but has a connection to
gauge fixing as well. 
To see this recall that the solution of Eq.~(\ref{eq:ScalarWaveGradDiv}) can be expressed via the solution of
the same equation with a simpler right-hand side. Indeed, transform Eq.~(\ref{eq:ScalarWaveGradDiv})
into the $(\bk,s)$-domain using the three-dimensional spatial Fourier and the 
one-dimensional temporal Laplace transforms:
\begin{align}
\label{eq:KSdomainWave}
-(\vert\bk\vert^{2}+\frac{s^{2}}{c^{2}})\tilde{\bE}(\bk,s)=
\frac{1}{s\varepsilon_{0}}\bk\bk\cdot\tilde{\bJ}(\bk,s)+s\mu_{0}\tilde{\bJ}(\bk,s).
\end{align}  
Solution of this algebraic equation can be written as
\begin{align}
\label{eq:KSDomainSolution}
\tilde{\bE}(\bk,s)=
-\frac{c^{2}}{s}\bk\bk\cdot\tilde{\bA}(\bk,s)-s\tilde{\bA}(\bk,s),
\end{align}  
where
\begin{align}
\label{eq:KSA}
\tilde{\bA}(\bk,s)=\frac{\mu_{0}}{\vert\bk\vert^{2}+s^{2}/c^{2}}\tilde{\bJ}(\bk,s),
\end{align}
and it is easy to deduce by inverse transformations that the new quantity $\bA(\bx,t)$
satisfies
\begin{align}
\label{eq:AEquation}
\Delta\bA(\bx,t)-\frac{1}{c^{2}}\partial_{t}^{2}\bA(\bx,t)=-\mu_{0}\bJ(\bx,t).
\end{align}
The inverse transformation of Eq.~(\ref{eq:KSDomainSolution}) shows that 
there is a (spatially) local relation between the solutions
of Eq.~(\ref{eq:ScalarWaveGradDiv}) and Eq.~(\ref{eq:AEquation}), namely,
\begin{align}
\label{eq:EviaA}
\bE(\bx,t)=c^{2}\nabla\nabla\cdot\int_{t_{0}}^{t}\bA(\bx,t')\,{\rm d}t'
-\partial_{t}\bA(\bx,t).
\end{align} 
On the other hand we know that a similar local relation exists between the fields and the vector and scalar
potentials, i.e.,
\begin{align}
\label{eq:FieldPotentials}
\bE(\bx,t)=-\nabla\phi(\bx,t)-\partial_{t}\bA(\bx,t).
\end{align} 
We see that the two expressions, (\ref{eq:EviaA}) and (\ref{eq:FieldPotentials}), are identical, 
if we employ the Lorenz gauge
\begin{align}
\label{eq:LorenzGauge}
\nabla\cdot\bA(\bx,t)+\frac{1}{c^{2}}\partial_{t}\phi(\bx,t)=0,
\end{align}
in the following time-integrated form:
\begin{align}
\label{eq:LorenzGaugeIntegrated}
c^2\nabla\cdot\int_{t_{0}}^{t}\bA(\bx,t')\,{\rm d}t'=-\phi(\bx,t).
\end{align}
The scalar potential, in general, also obeys a wave equation 
\begin{align}
\label{eq:ScalarPotential}
\Delta\phi(\bx,t)-\frac{1}{c^{2}}\partial_{t}^{2}\phi(\bx,t)=-\frac{1}{\varepsilon_{0}}\rho(\bx,t),
\end{align}
which can be written as
\begin{align}
\label{eq:ScalarPotentialCurrent}
\Delta\phi(\bx,t)-\frac{1}{c^{2}}\partial_{t}^{2}\phi(\bx,t)=\frac{1}{\varepsilon_{0}}\nabla\cdot\int_{t_{0}}^{t}\bJ(\bx,t')\,{\rm d}t',
\end{align}
if we use the following time-integrated form of the continuity equation:
\begin{align}
\label{eq:Continuity}
\nabla\cdot\int_{t_{0}}^{t}\bJ(\bx,t')\,{\rm d}t'=-\rho(\bx,t).
\end{align}
Thus, employing the compatibility relation (\ref{eq:IntCompatibility}) to derive 
the wave equation and 
the Lorenz gauge fixing (\ref{eq:LorenzGaugeIntegrated}) required for
the unique representation of fields via potentials are two completely equivalent 
procedures. Moreover, we have also established that this amounts to (follows from) relating the sources of the 
scalar and vector potentials via the integrated continuity 
equation (\ref{eq:Continuity}). The underlying mathematical assumption
here is that the current density $\bJ$ accounts
for {\it all} the sources of the electromagnetic field. Hence, Eq.~(\ref{eq:Continuity})
describes all possible variations of the charge density $\rho$. From the physical point of view,
we assume that any eventual ``bare'' charge can only be created by some explicit dynamical process
which disrupts the initial electrical neutrality 
of the substance (e.g. chemical reaction). Therefore, a current --
motion of charges -- precedes and causes the accumulation of  
charge. This is an often neglected and more subtle part of the general
causality assumption behind the radiation formula. It treats static field as a
stationary long-time limit of an initially transient electromagnetic field. The very
existence of this limit and therefore static fields as such is thus an open problem
in this formulation.

\section{Time-domain radiation formula}
A detailed derivation of the radiation formula presented below can be found in De~Hoop (1995). 
It proceeds by computing the inverse
Fourier transform of the $(\bk,s)$-domain solution (\ref{eq:KSDomainSolution}). 
We arrive at
\begin{align}
\label{eq:XSDomainSolution}
\hat{\bE}(\bx,s)=\nabla\nabla\cdot\frac{c^{2}}{s}\hat{\bA}(\bx,s)-s\hat{\bA}(\bx,s),
\end{align}  
with
\begin{align}
\label{eq:AXSDomain}
\hat{\bA}(\bx,s)=\mu_{0}\int_{\bx'\in{\mathbb R}^{3}}g(\bx-\bx',s)\hat{\bJ}(\bx',s)\,{\rm d}\bx',
\end{align}
where the scalar Green's function is
\begin{align}
\label{eq:GreenFunction}
g(\bx,s)=\frac{e^{-(s/c)\vert\bx\vert}}{4\pi\vert\bx\vert}.
\end{align}
At this stage it is convenient to carry out the spatial differentiations in
the first term on the right in Eq.~(\ref{eq:XSDomainSolution}). The resulting formula is
\begin{align}
\label{eq:XSTensorFormula}
\begin{split}
\hat{\bE}(\bx,s)&=\int_{\bx'\in{\mathbb R}^{3}}
\frac{e^{-(s/c)\vert\bx-\bx'\vert}}{4\pi\vert\bx-\bx'\vert^{3}}
\left[3{\mathbb Q}-{\mathbb I}\right]\frac{1}{s\varepsilon_{0}}\hat{\bJ}(\bx',s)
\,{\rm d}\bx'
\\
&+\int_{\bx'\in{\mathbb R}^{3}}
\frac{e^{-(s/c)\vert\bx-\bx'\vert}}{4\pi\vert\bx-\bx'\vert^{2}}
\left[3{\mathbb Q}-{\mathbb I}\right]\frac{1}{\varepsilon_{0}c}\hat{\bJ}(\bx',s)
\,{\rm d}\bx'
\\
&+\int_{\bx'\in{\mathbb R}^{3}}
\frac{e^{-(s/c)\vert\bx-\bx'\vert}}{4\pi\vert\bx-\bx'\vert}
\left[{\mathbb Q}-{\mathbb I}\right]\frac{s}{\varepsilon_{0}c^{2}}\hat{\bJ}(\bx',s)
\,{\rm d}\bx'.
\end{split}
\end{align}
The three terms above are called near-, intermediate-, and far-field contributions 
in accordance with their spatial decay factors. The tensors act as
\begin{align}
\label{eq:QTensor}
{\mathbb Q}\bJ(\bx')&=\frac{(\bx-\bx')}{\vert\bx-\bx'\vert}
\left(\frac{(\bx-\bx')}{\vert\bx-\bx'\vert}\cdot\bJ(\bx')\right),
\\
\label{eq:ITensor}
{\mathbb I}\bJ(\bx')&=\bJ(\bx').
\end{align}
The magnetic field can be found from a similar formula
\begin{align}
\label{eq:XSTensorFormulaH}
\begin{split}
\hat{\bH}(\bx,s)&=-\int_{\bx'\in{\mathbb R}^{3}}
\frac{e^{-(s/c)\vert\bx-\bx'\vert}}{4\pi\vert\bx-\bx'\vert^{2}}
\bTheta\times\hat{\bJ}(\bx',s)
\,{\rm d}\bx'
\\
&-\int_{\bx'\in{\mathbb R}^{3}}
\frac{e^{-(s/c)\vert\bx-\bx'\vert}}{4\pi\vert\bx-\bx'\vert}
\frac{s}{c}\bTheta\times\hat{\bJ}(\bx',s)
\,{\rm d}\bx',
\end{split}
\end{align}
where 
\begin{align}
\label{eq:ThetaVec}
\bTheta=\frac{\bx-\bx'}{\vert\bx-\bx'\vert}.
\end{align}
Finally, the inverse Laplace transform gives the explicit time-domain formulas 
\begin{align}
\label{eq:XTFormula}
\begin{split}
\bE(\bx,t)&=\int_{\bx'\in{\mathbb R}^{3}}
\frac{3{\mathbb Q}-{\mathbb I}}{4\pi\vert\bx-\bx'\vert^{3}}
\frac{1}{\varepsilon_{0}}\int_{t_{0}}^{t_{\rm R}}\bJ(\bx',t')\,{\rm d}t'
\,{\rm d}\bx'
\\
&+\int_{\bx'\in{\mathbb R}^{3}}
\frac{3{\mathbb Q}-{\mathbb I}}{4\pi\vert\bx-\bx'\vert^{2}}
\frac{1}{\varepsilon_{0}c}\bJ(\bx',t_{\rm R})
\,{\rm d}\bx'
\\
&+\int_{\bx'\in{\mathbb R}^{3}}
\frac{{\mathbb Q}-{\mathbb I}}{4\pi\vert\bx-\bx'\vert}
\frac{1}{\varepsilon_{0}c^{2}}\partial_{t}\bJ(\bx',t_{\rm R})
\,{\rm d}\bx'
\end{split}
\end{align}
and
\begin{align}
\label{eq:XTFormulaH}
\begin{split}
\bH(\bx,t)&=-\int_{\bx'\in{\mathbb R}^{3}}
\frac{1}{4\pi\vert\bx-\bx'\vert^{2}}
\bTheta\times\bJ(\bx',t_{\rm R})
\,{\rm d}\bx'
\\
&-\int_{\bx'\in{\mathbb R}^{3}}
\frac{1}{4\pi\vert\bx-\bx'\vert}
\bTheta\times\frac{1}{c}\partial_{t}\bJ(\bx',t_{\rm R}),
\,{\rm d}\bx'
\end{split}
\end{align}
where for the first time we meet the retarded time
\begin{align}
\label{eq:RetardetTime}
t_{\rm R}=t-\frac{\vert\bx-\bx'\vert}{c},
\end{align}
which is an analogue of the $\pm ct$ shift of the d'Alembert solution (\ref{eq:Dalamber}).
The spatial integration in (\ref{eq:XTFormula}) will in practice be limited
to a finite domain $D_{\rm s}$ occupied by the source. 

\section{Explanation of near-field superluminal velocities}
With the radiation formula (\ref{eq:XTFormula}) at hand we can start analyzing
the shape of the waveforms received at some distance from the source.
A waveform measured by the dipole antenna of the type shown in Fig.~\ref{fig:ExperimentalSetup}
gives the time-evolution of a single component of the electric
field strength determined by the orientation of the dipole and modified
by the the antenna-cable-oscilloscope receiving tract.
The latter modification, however, is the same for all receiver
locations, i.e., does not depend on the distance from the source. 
Only in the close proximity of the source the expected small multiple reflections
between the two antennas may slightly modify the tail of the waveform with respect 
to what is predicted by Eq.~(\ref{eq:XTFormula}).

A simplified formula for small source and receiver dipoles,
parallel to each other and located in a plane orthogonal to their orientation 
(as in Fig.~\ref{fig:ExperimentalSetup}) is
\begin{align}
\label{eq:DipolesFormula}
V(\bx,t)\sim -\frac{1}{4\pi\varepsilon_{0} R^{3}}
\int_{t_{0}}^{t_{\rm R}}I(t')\,{\rm d}t'
-\frac{1}{4\pi\varepsilon_{0}c R^{2}}
I(t_{\rm R})
-\frac{1}{4\pi\varepsilon_{0}c^{2} R}
\partial_{t}I(t_{\rm R}),
\end{align}
where $R$ is the distance between the dipoles, $t_{\rm R}=t-R/c$, $V$ is the
measured waveform, and $I$ is the signal (current) in the source antenna. 

\begin{figure}[t]
	\includegraphics[width=\textwidth]{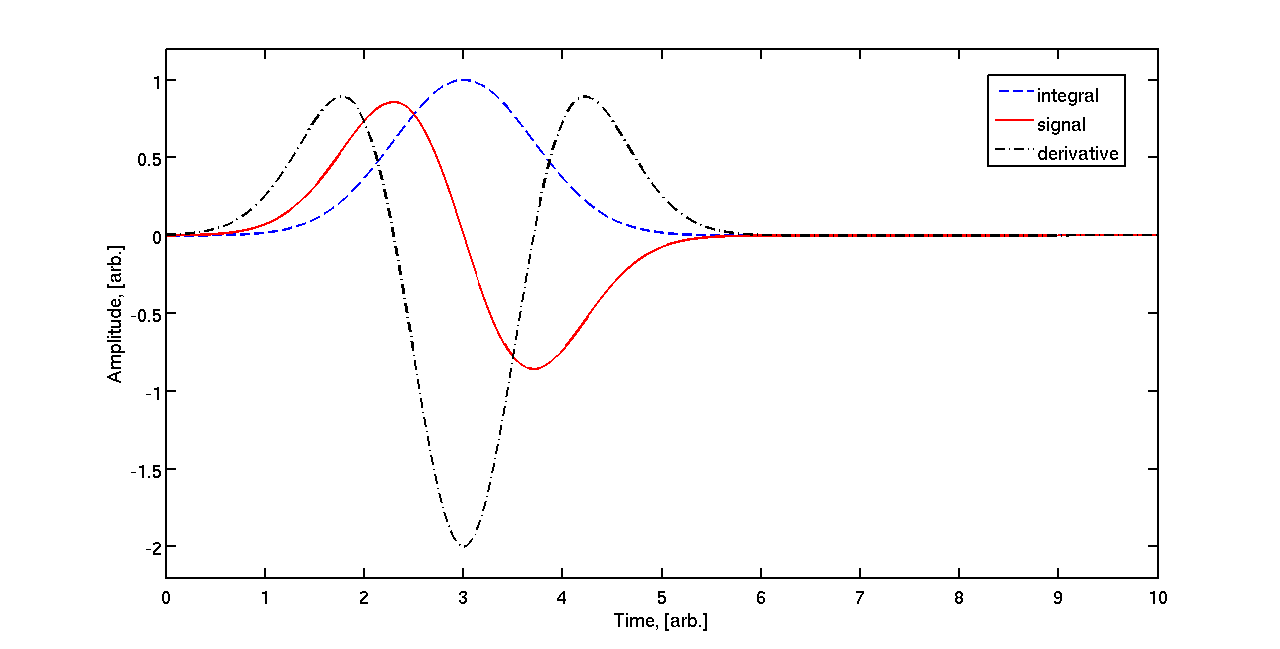}
	\captionof{figure}{Back-in-time shifting of extrema. Depicted are the signal, its
                           cumulative time-integral, and derivative. The
                           first extremum of the integral is further (later) 
                           in time, whereas the first extremum of the derivative 
                           is earlier in time than the first extremum of the 
                           original signal.}
	\label{fig:timeshift}
\end{figure}

Obviously, the received waveform is a function not only of the signal $I$ at an
earlier time, but also of its time-derivative and time-integral. 
Figure~\ref{fig:timeshift} shows the three functions in question
for the current of the form
\begin{align}
\label{eq:Current}
I(t)=-2\alpha(t-t_{\rm c})e^{-\alpha(t-t_{\rm c})^{2}},
\end{align}
whose integral and derivative can be computed analytically. As can be seen from
Fig.~\ref{fig:timeshift}
the extrema of the three functions are time-shifted with
respect to each other. In the present case we can talk about the first extremum
only, since the integral does not have a second one.
With respect to this first extremum we notice that the integral has it later, 
and the derivative has it earlier in time than the function itself. 
This is a purely mathematical phenomenon and it will be observed with any function.

The received waveform is a weighted sum of these three functions. The weight
coefficients are the functions of distance $R$ between the source and the receiver,
where the term with $R^{-3}$ will be dominant close to the source (near-field zone), and the
term with $R^{-1}$ is dominant for large $R$ (far-field zone). This means that
in the near-field zone the received waveform will look almost like the integral,
and in the far-field zone it will look almost like the derivative of the current.

The two-point procedure described in Section~2 deduces the velocity of the
electromagnetic pulse from the waveform measured at two locations where one is
further away from the source. Thus, according to Eq.~(\ref{eq:Current}) the values
of the weighting factors will be different for these two locations. Not only
the waveform will be smaller in amplitude for the more distant of the two locations,
but the relative weight of the three contributions will change as well. In general
the relative weight of the last term increases, whereas the relative weight of the first term decreases
with distance. Hence, we may expect the shape of the waveform to gradually change from
the one dominated by the time-integrated signal to the one dominated by the
time-derivative of the signal. As can be seen from Fig.~\ref{fig:timeshift}
this means a shift of an extremum towards earlier times. Such backward in time
``motion'' will be superimposed on the normal time-delay which shifts the 
received waveform to the right along the time axis as one moves further
away from the source.

The latter observation explains our failure to measure the speed of light in
Section~2. Indeed, the first measurement location in Fig.~\ref{fig:fastnear}
is close enough to the source for the influence of the near- and intermediate-field 
terms to be significant. Therefore, the overall waveform is somewhat shifted to the
right along the time-axis relative to the time-delayed derivative of the original signal,
which is dominant at larger distances.   
At the second location the influence of the near- and intermediate-field terms is smaller
and so is the relative shift. Let us express the arrival times of the measured extrema
at two locations as
\begin{align}
\label{eq:ArrivalTimes}
\begin{split}
t_{\rm arr}(R_{1})&=t_{0}+\frac{R_{1}}{c}+\Delta t(R_{1}),
\\
t_{\rm arr}(R_{2})&=t_{0}+\frac{R_{2}}{c}+\Delta t(R_{2}),
\end{split}
\end{align} 
where $\Delta t(R)$ is the relative shift due to the near- and 
intermediate-field terms. We give it a positive sign to emphasize that
the relative time-shift is happening towards later (positive) times
on top of the normal propagation induced positive
time-delay $R/c$. 

The speed of light was deduced in Section~2 from the
following simple two-point calculation:
\begin{align}
\label{eq:SpeedCalculation}
c_{\rm exp}=\frac{R_{2}-R_{1}}{t_{\rm arr}(R_{2})-t_{\rm arr}(R_{1})},
\end{align}
applied to the extrema of the measured waveform.
It is easy to see, that without the additional relative time-shifts $\Delta t$,
substitution of (\ref{eq:ArrivalTimes}) would have given us the exact speed of light.
Now, however, the denominator in (\ref{eq:SpeedCalculation}) is
\begin{align}
\label{eq:TimeDifference}
t_{\rm arr}(R_{2})-t_{\rm arr}(R_{1})=\frac{R_{2}-R_{1}}{c}+\Delta t(R_{2})-\Delta t(R_{1}).
\end{align}
Since the relative time-shift $\Delta t(R)$ is diminishing with distance, 
we have in general
\begin{align}
\label{eq:DeltaT}
\Delta t(R_{2})-\Delta t(R_{1})<0, \;\;\; R_{2}>R_{1},
\end{align} 
Thus, the measured time difference is always smaller then the expected one. If this
time difference is positive we get
\begin{align}
\label{eq:MesSpeed}
c_{\rm exp}>c,
\end{align}
i.e., the measured speed is superluminal -- exactly as we have observed.

That this superluminal behavior is not visible in Fig.~\ref{fig:normalfar},
and in the majority of experiments with light, can also be explained. The 
near- and intermediate-field terms, which cause the additional time shift, decay not only 
relatively with respect to the far-field term, but also absolutely, i.e.,
their influence is practically not measurable in the far field. The far-field zone is
a notion relative to the wavelength. Usually, the far-field zone starts
beyond a few tens of wavelengths, i.e., is very close to the source for the 
visual light. 

The measured speed reflects two competing
phenomena -- outward propagation and the diminishing additional positive time shift.  
In the far-field zone the first of these phenomena gradually takes over, and
we may expect the local two-point measurement procedure to yield values
of speed progressively approaching the speed of light from above, as
if the pulse was decelerating. In the near-field zone, however, the
additional time shift is large enough to be detectable. We saw how it
yields superluminal values for the pulse speed. Obviously, we need to let go
of the idea of a constant speed and consider a more general concept of 
a local velocity which is a function of distance from the source.
For example, we could introduce a local velocity as a limit of the two-point measurement 
procedure in the neighborhood of some location $R$ as
\begin{align}
\label{eq:LocalDefinition}
v_{\rm loc}(R)=\lim\limits_{\epsilon\rightarrow 0}
\frac{\epsilon}{t^{\rm ex}(R+\epsilon)-t^{\rm ex}(R)},
\end{align}
where $t^{\rm ex}(R)$ is the time of the waveform extremum measured at location $R$.
This definition, however, is hard to apply as we do not know the above extrema 
times explicitly. 

\section{Locally negative velocities}
\begin{figure}[t]
	\includegraphics[width=\textwidth]{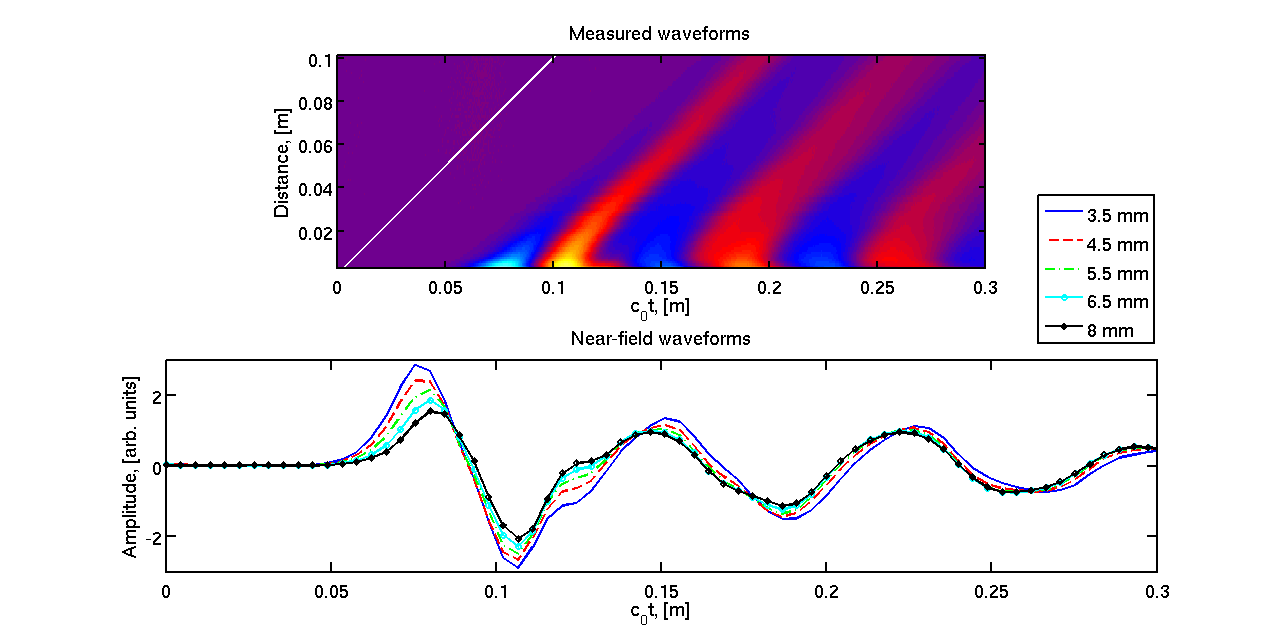}
	\captionof{figure}{Experimental demonstration of local negative pulse 
velocity in the near-field of a small dipole antenna. Upper image shows the space-time
dynamics of the signal. Close to the source (bottom of the image) the wave makes a small
bend to the left along the time axis -- negative velocity region. Further from the source
the wave dynamics approaches the light-cone behavior (white line). Lower plot shows the wave
dynamics in the near-field zone. While the wavefront and the first slope shift towards later 
times (positive velocity), the extrema and the whole inner part of the waveform shift to 
the left (negative velocity).}
	\label{fig:negative}
\end{figure}

As was mentioned above superluminal pulse propagation 
is observed, if the measured time difference in the two-point procedure is smaller 
than the expected $R/c$, but remains {\it positive}. If this difference is negative, 
then the measured velocity would also become negative. Are such negative velocities 
possible, and what does it mean?
The answer can be obtained with the formulas and understanding we have acquired
in the previous section. Indeed, a negative velocity could be measured in a two-point
procedure, if the anomalous positive time-shift we discussed was disappearing faster 
with $R$ than the normal growth of the positive time-shift due to propagation, i.e.,
\begin{align}
\label{eq:NegativeTimeDiff}
\frac{R_{2}-R_{1}}{c}<\Delta t(R_{1})-\Delta t(R_{2}).
\end{align}      
Recently an experiment was reported by the author, Budko (2009). 
It showed that this effect is indeed present and can be observed with the setup 
which we used above to demonstrate the superluminal pulse velocities. 
Very close to the source the waveform shows a small but detectable movement
towards earlier times (to the left on the scope screen) as we move the receiver away
from the source in small steps. Figure~\ref{fig:negative} shows the results of this experiment. 
The measurements had to be carried out very close to the source. For the $75$-ps pulse
that was used in the experiment the negative velocity region is within $10$~mm distance from the 
source. The extent of the negative-velocity region was initially 
predicted by a simulation similar to the one considered in the next section.
The actual measurements also contain multiple reflections
between the source and receiver dipoles. These reflections show as a small bump
in the slope of the waveform between $c_{0}t=0.1$ and $c_{0}t=0.15$~m. This was verified 
by inserting a larger metallic scatterer in the plain of the source and observing 
an increase of the said bump as well as the appearance of multiples further down the 
tail of the waveform. 
Apart from that the leftward shift of the inner part of the waveform is clearly visible.  
The upper image in Fig.~\ref{fig:negative} is a rotated version of a typical light-cone diagram from special 
relativity. In fact, the white line indicates the way the light cone is supposed to
be. If the light was indeed following its ``cone'', then the colored stripes in the
image should all follow that white line. This happens only further away from the source.
Whereas in the bottom of the image there is an obvious near-field distortion -- 
a leftward bend corresponding to the negative velocity of propagation. 

Such behavior is very counterintuitive and may even seem anti-causal. Indeed, it looks like 
the signal reaches a more distant observer faster than a closer one. 
Although information may be and often is encoded in extrema of a waveform the speed at which it 
reaches the observer is hard to define. We notice from the experimental data in Figure~\ref{fig:negative}
that the front of the wave travels with a positive velocity. Moreover, 
for a source current with a sharp temporal boundary $I(t)=0$, $t\le t_{0}$ we can 
prove analytically that the wavefront always travels outwards at the speed of light,
since near- and intermediate-field terms are exactly zero for $t=t_{\rm R}$ in that case. 
For the same reason the end of the waveform would also travel
outwards and at the speed of light. Thus, the beginning and the end of an information
carrying wave-packet would arrive at two observes in a proper relativistically consistent 
temporal order, i.e.,
later for the more distant observer. However, due to the anomalous temporal shift
of extrema the more distant observer may actually receive the information sooner than the
closer one. Of course, we are talking about the near-field zone and two very close observers,
but nevertheless. Explanation seems to be in the fact that the two observes are
causally related to the source of the field, but not to each other. The situation becomes
less paradoxical if we put it this way: it is not that
the more distant observer gets the information {\it faster} than normal. No, the beginning 
of the information packet arrives at the speed of light. It is that the closer of the
two observers receives the same information {\it slower} than normal, due to the additional
positive time-shift of the waveform in the near-field zone. The same explanation
helps to comprehend the superluminal impulse velocities observed earlier.

\section{Negative power flow}
At the moment the strange velocities we observe in the near-field zone seem to be 
tightly linked to the two-point
measurement procedure and our choice of extrema as the reference points. 
To show that the two-point procedure is not really essential here let us look at 
the power flow in the near-field zone. An analysis along the 
following lines, although focused on the Hertzian dipole with an initial static field, 
can be found in Schantz (2001). The instanteneous
power density flowing through location $\bx$ at time $t$
is given by the Poynting vector
\begin{align}
\label{eq:Power}
\bS(\bx,t)=\bE(\bx,t)\times\bH(\bx,t).
\end{align}
In our particular experimental arrangement with a small source the current 
density can be approximated by 
\begin{align}
\label{eq:DeltaSource}
\bJ(\bx,t)=\bd I(t)\delta(\bx-\bx_{\rm s}),
\end{align}
where $\bd$ is a unit vector in the direction of the current.
For the geometry depicted in Fig.~\ref{fig:ExperimentalSetup} we have
\begin{align}
\label{eq:PowerParallel}
\bS(R,t)=P(R,t)\bTheta,
\end{align}
where the unit vector $\bTheta$ points away from the source towards the receiver 
and the power density amplitude is
\begin{align}
\label{eq:PowerMagnitude}
\begin{split}
P(R,t)&=\left[\frac{1}{4\pi\varepsilon_{0} R^{3}}
\int_{t_{0}}^{t_{\rm R}}I(t')\,{\rm d}t'
+\frac{1}{4\pi\varepsilon_{0}c R^{2}}I(t_{\rm R})
+\frac{1}{4\pi\varepsilon_{0}c^{2} R}
\partial_{t}I(t_{\rm R})\right]\times
\\
&\left[\frac{1}{4\pi R^{2}}I(t_{\rm R})
+\frac{1}{4\pi c R}
\partial_{t}I(t_{\rm R})\right]
\\
&=\frac{1}{(4\pi)^{2}\varepsilon_{0}R^{5}}I(t_{\rm R})\int_{t_{0}}^{t_{\rm R}}I(t')\,{\rm d}t'
+
\frac{1}{(4\pi)^{2}\varepsilon_{0}c R^{3}}\partial_{t}I(t_{\rm R})\int_{t_{0}}^{t_{\rm R}}I(t')\,{\rm d}t'+
\\
&\left[\frac{1}{4\pi\varepsilon_{0}^{1/2}c^{1/2} R^{2}}I(t_{\rm R})+
\frac{1}{4\pi\varepsilon_{0}^{1/2}c^{3/2}R}\partial_{t}I(t_{\rm R})\right]^{2}.
\end{split}
\end{align}
While the last term is undoubtedly positive at all times, the remaining two terms 
may, in principle, become negative due to the mixed products of the current, its
derivative, and integral, which as we know can all have different signs at the same time
-- see Fig.~\ref{fig:timeshift}. If a negative power flow can be detected, 
the phenomenon would again be
confined to the near-field zone, as the dominant far-field term is positive.

\begin{figure}[t]
	\includegraphics[width=\textwidth]{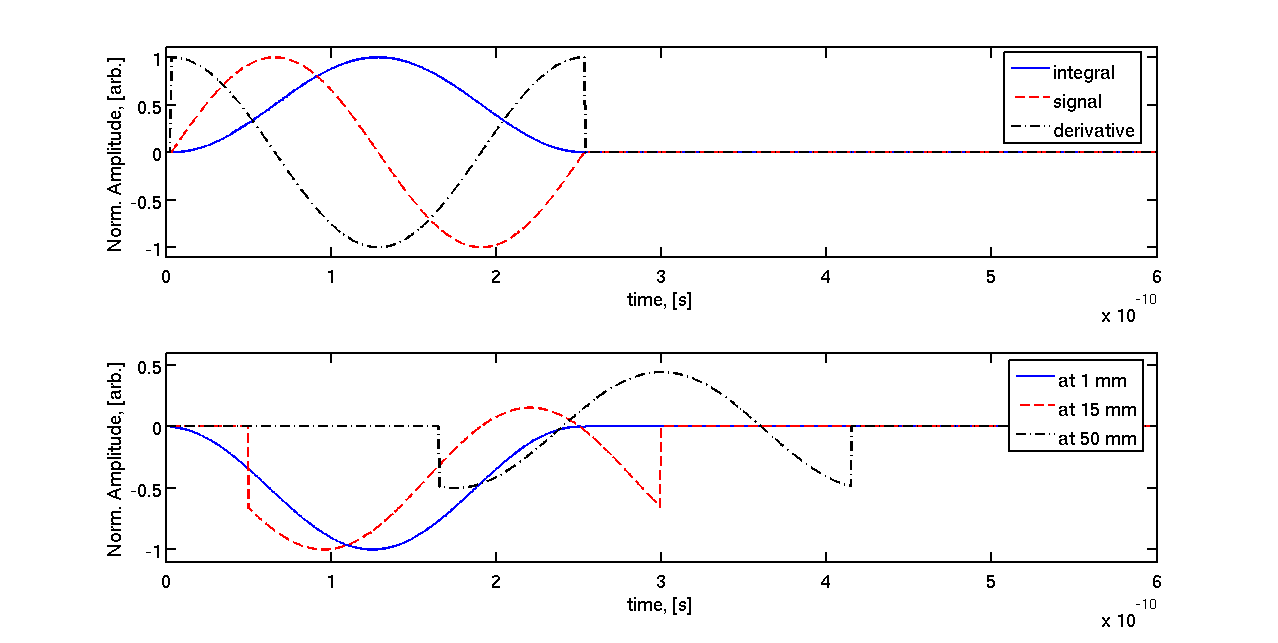}
	\captionof{figure}{Top: mono-cycle source current, its integral, and derivative (normalized).
                           Bottom: Simulated electric field at various distances from the source (normalized).
                           Close to the source the received signal resembles the integral taken with a 
                           negative sign. Further it deforms into a shifted negative of the signal, and finally transforms into 
                           a negative of the derivative of the signal. The negative velocity phenomenon is obvious between $1$~mm 
                           and $15$~mm distances -- waveform shifts leftwards.}
	\label{fig:products}
\end{figure}

Let us consider another current pulse with analytically known derivative and integral.
This time, however, we shall look at a pulse called 
{\it monocycle} with well-defined beginning and end:
\begin{align}
\label{eq:Monocycle}
\begin{split}
I(t)=\left\{
\begin{array}{l}
\sin(\omega t),\;\;\;t\in[t_{0},t_{0}+2\pi/\omega],\\
0,\;\;\;t\notin[t_{0},t_{0}+2\pi/\omega],
\end{array}
\right.
\end{split}
\end{align}
where $\omega$ is the carrier frequency.
The derivative and integral of this signal within the time interval $[t_{0},t_{0}+2\pi/\omega]$
are:
\begin{align}
\label{eq:DerIntMono}
\begin{split}
\partial_{t}I(t)&=\omega\cos(\omega t),
\\
\int_{t_{0}}^{t}I(t')\,{\rm d}t'&=\frac{1}{\omega}\left[\cos(\omega t_{0})-\cos(\omega t)\right],
\end{split}
\end{align}
and are zero outside the said interval. The normalized versions of these functions 
are shown in the top plot of Fig.~\ref{fig:products} for $\omega=8\pi\times 10^{9}$, i.e.,
for a $4$~GHz carrier frequency.
The bottom plot of that figure shows the results of simulation of the electric field at three
different distances from the source: $1$, $15$, and $50$~mm. We use Eq.~(\ref{eq:DipolesFormula})
to simulate the waveforms. 
All pertaining deformations of the waveform are now clearly visible,
as it resembles the integral of the signal in the near-field zone and gradually
transforms into the derivative of the signal as we measure further away. We also
see the negative velocity phenomenon as the inner part of the waveform at $15$~mm
is clearly shifted to the left, instead of being shifted to the right of the waveform
corresponding to $1$~mm distance. At the same time the wavefront and the end of the waveform 
do shift rightwards as predicted. 

\begin{figure}[t]
	\includegraphics[width=\textwidth]{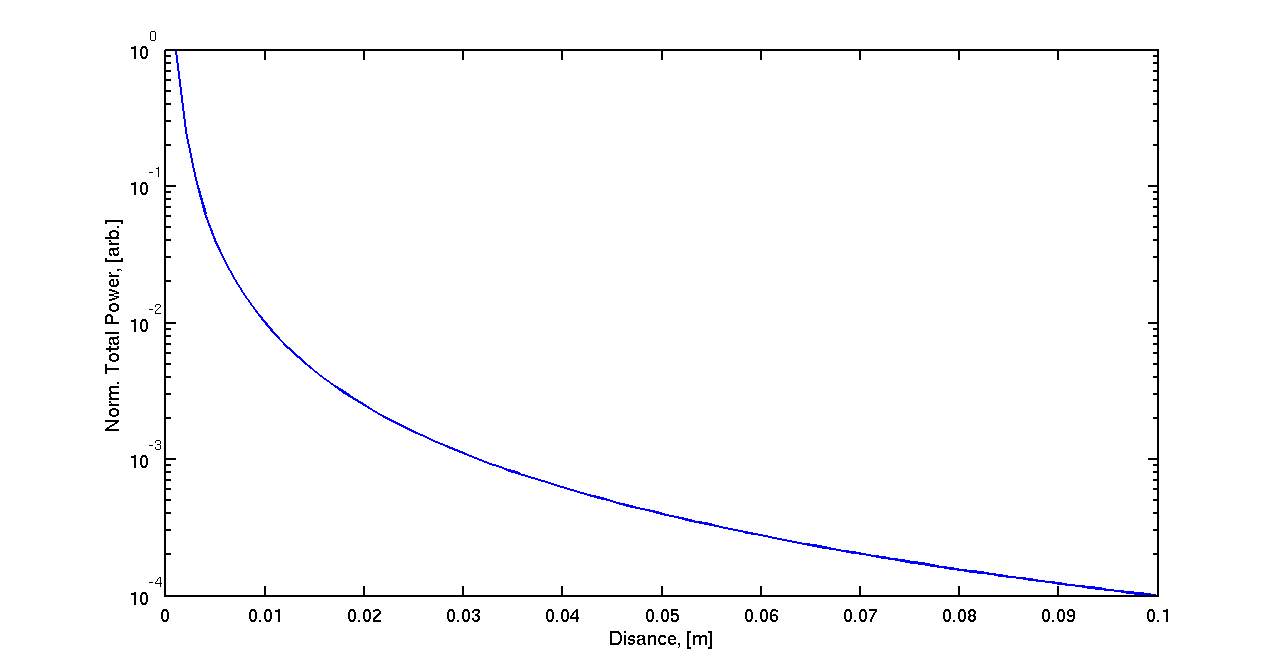}
	\captionof{figure}{Normalized time-integrated power as a function of 
distance from the source. This quantity is positive everywhere.}
	\label{fig:intpower}
\end{figure}

\begin{figure}[t]
	\includegraphics[width=\textwidth]{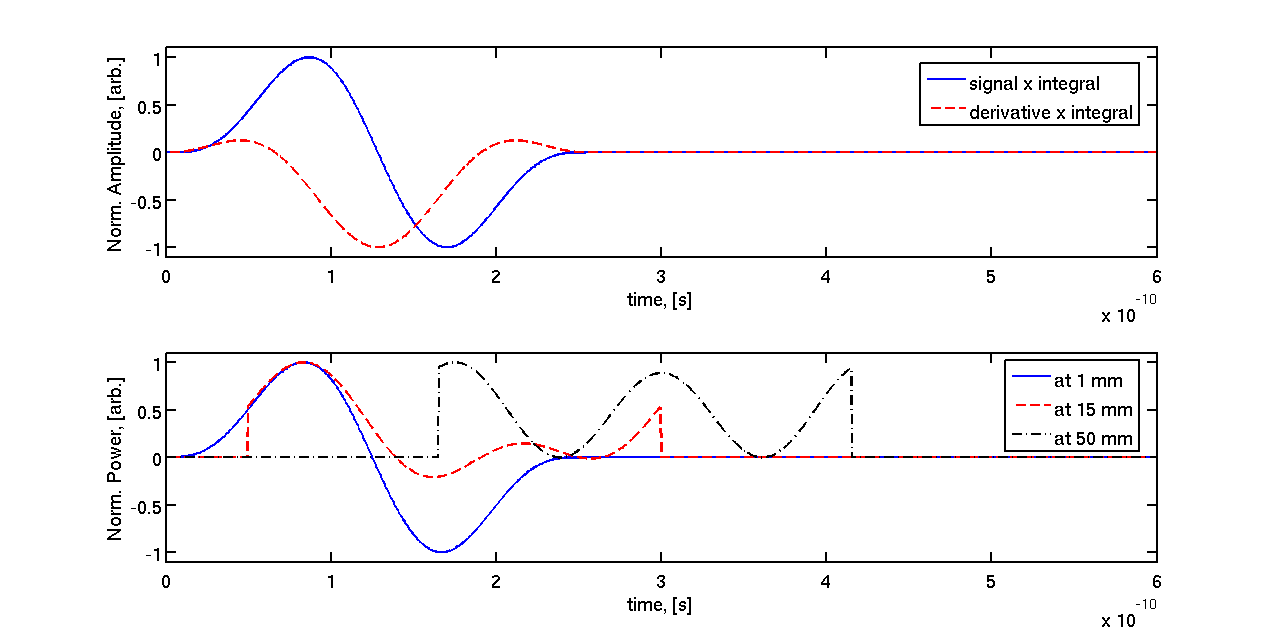}
	\captionof{figure}{Top: normalized products of the signal and its integral, and derivative
of the signal and signal's integral corresponding to the source current shown in 
Fig.~\ref{fig:products}. Notice the presence of negative parts in both cases. 
Bottom: normalized instantaneous power flow at various distances from the source. In the
near-field zone the power flow can become negative for some time. During that time
power flows towards the source. 
Further from the source the power flow looks positive at all times (see, however, Fig.~\ref{fig:energy}).}
	\label{fig:power}
\end{figure}

To understand, at least mathematically, the emergence of the negative power flow consider a very small distance $R$, 
such that the first terms in Eq.~(\ref{eq:PowerMagnitude}) are dominant. These terms contain mixed products 
of the current, its integral, and derivative. For the particular current given by Eq.~(\ref{eq:Monocycle})--(\ref{eq:DerIntMono})
we see that these mixed products will be negative over a certain period of the time
-- see Fig.~\ref{fig:power} (top). Although we do not prove it here rigorously for 
an arbitrary current, computer simulations and common sense indicate that the time-integrated 
power is positive at all distances, including the near-field zone -- see Fig.~\ref{fig:intpower}. 
This, however, does
not exclude the possibility of instantaneous power being negative for some time. 

In Figure~\ref{fig:power} (bottom) we see the power $P(R,t)$ as a function of time
at three distances from the source. The negative power flow
is clearly visible in this plot. As we move towards the far-field
zone the power flow {\it looks} positive at all times. This, however,
is not true as we shall see in the next Section.  

Power flow tells us how much energy is coming from a certain direction 
in a unit of time. If a
power flow is negative it means that the energy is decreasing with time,
or that the power is flowing towards the source of radiation, instead of 
away from it. Since the Poynting vector is related to the density of the
linear momentum of the electromagnetic field, we may expect the instantaneous
pressure of light to change its direction over extremely short time intervals 
in the near-field zone. Not only the light will be pushing away from the source,
as it is usually the case, but pulling towards the source as well. 
This may induce a very rapid mechanical oscillation 
on a neutral test particle -- an 
opto-mechanical effect which might be of interest in nano-physics.

\section{Energy velocity}
We can formally relate the power flow and the energy density in our experimental
setup via the following formula:
\begin{align}
\label{eq:PowerEnergy}
P(\bx,t)\bTheta=v_{\rm e}(\bx,t){\mathcal E}(\bx,t)\bTheta,
\end{align} 
where ${\mathcal E}$ is the local instantaneous energy of the electromagnetic field.
From where the energy velocity along $\bTheta$ is expressed as
\begin{align}
\label{eq:EnergyVelocity}
v_{\rm e}(\bx,t)=\frac{P(\bx,t)}{{\mathcal E}(\bx,t)}.
\end{align}
Simulations of the energy velocity for a transient Hertzian dipole were presented by
Marcano~\&~Diaz (2006). Since, as we saw above, the local power flow can be negative, while the
field energy is always positive, the near-field zone results of Marcano~\&~Diaz (2006)
naturally contained locally negative energy velocities.   

\begin{figure}[t]
	\includegraphics[width=\textwidth]{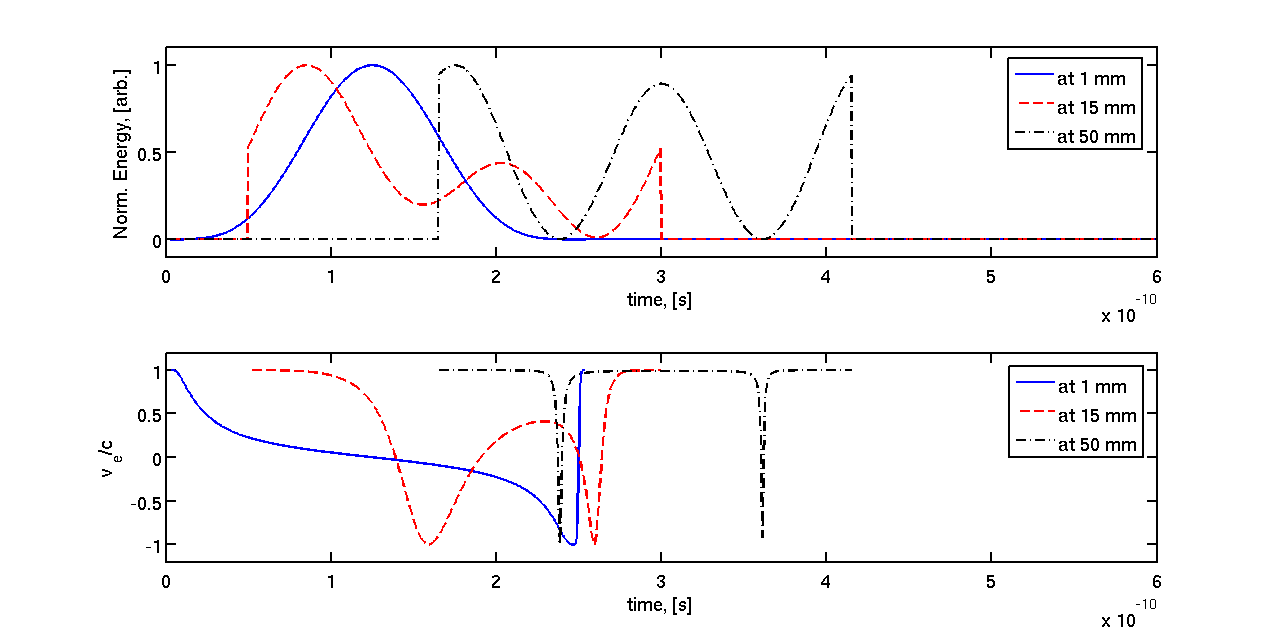}
	\captionof{figure}{Top: normalized local instantaneous energy at various distances 
corresponding to the source current shown in 
Fig.~\ref{fig:products}.  
Bottom: Relative energy velocity, $v_{\rm e}/c$, 
at various distances from the source.}
	\label{fig:energy}
\end{figure}

The power amplitude for our measurement configuration is given by Eq.~(\ref{eq:PowerMagnitude}), whereas 
the energy density at location $R$ as a function of time is
\begin{align}
\label{eq:EnergyDensity}
\begin{split}
{\mathcal E}(R,t)&=\varepsilon_{0}\left[\frac{1}{4\pi\varepsilon_{0} R^{3}}
\int_{t_{0}}^{t_{\rm R}}I(t')\,{\rm d}t'
+\frac{1}{4\pi\varepsilon_{0}c R^{2}}I(t_{\rm R})
+\frac{1}{4\pi\varepsilon_{0}c^{2} R}
\partial_{t}I(t_{\rm R})\right]^{2}+
\\
&\mu_{0}\left[\frac{1}{4\pi R^{2}}I(t_{\rm R})
+\frac{1}{4\pi c R}
\partial_{t}I(t_{\rm R})\right]^{2}.
\end{split}
\end{align}
This function for the monocycle source current Eq.~(\ref{eq:Monocycle}) 
and the corresponding instantaneous 
relative energy velocity $v_{\rm e}(R,t)/c$ are shown in Fig.~\ref{fig:energy}.
Similar to the waveform of the electric field, the waveform of the instantaneous 
energy deforms as one moves through the near-, intermediate-, and far-field 
zones. There is no doubt about the positivity of energy density at all locations
and times due to the intrinsic mathematical form of expression 
Eq.~(\ref{eq:EnergyDensity}). Thus, the negative values of the energy velocity
visible in Fig.~\ref{fig:energy} (bottom) are caused by the locally negative power 
amplitude. It is obvious, that the energy velocity varies between $-c$ and $c$.
It is also obvious that the energy velocities of the front and the end of the 
wave-packet are both exactly the speed of light, and that the deviations concern
the inner part of the waveform only. We notice that the subluminal and negative 
velocities are mainly featured in the near-field zone and that the far-field
energy waveform moves mostly at the speed of light. However, at the time instant
where both energy and power tend to zero, the energy velocity makes a sharp
swing to $-c$ and back even in the far-field zone. This shows that the power 
amplitude must be going slightly negative at the minima of the far-field
waveform, something we could not previously detect in Fig.~\ref{fig:power}.
Thus, although the local instantaneous energy velocity seems to be a very natural
concept, and its magnitude is explicitly bounded by the speed of light,
it also magnifies the effect of the first two terms in Eq.~(\ref{eq:PowerMagnitude}),
so that it is felt even in the far-field zone, albeit for a vanishingly small time. 
Whether this swing in the power flow can be detected remains to be seen.

\section{Conclusions}
We have discussed a series of unusual transient effects in the
electromagnetic radiation from a localized source. These effects demonstrate that the space-time
evolution of the electromagnetic field is nothing like a simple outwards
propagation of a spherical wave at the speed of light. In fact, the speed of 
light could only be assigned to the boundary between the field and the 
area free from it -- the wavefront and the end of a wave-packet. The inner
part of the wave-packet undergoes significant deformation in the course
of propagation. This deformation leads, in particular, to superluminal
and negative velocity results with the classical two-point velocity measurement 
procedure. We have uncovered that the source of this deformation is an
anomalous time delay, which the waveform undergoes in the immediate vicinity
of the source. It takes more time for, say, an extremum to build up close to the source
than a little further away. Hence, paradoxically, it takes more time for the information to
arrive at a closer observer than at an observer further away from the source.
Yet, both transmission times are within what is allowed by the relativity theory.    
While this effect can be observed only very close to the source, the other related
phenomenon can be measured in the intermediate-field zone as well.

The anomalous time delay rapidly decays with the distance from the source. Hence,
the time-interval measured in the two-point procedure is usually smaller than
what is expected from the propagation at the speed of light. This explains
the observed superluminal results and shows the necessity of near-field corrections
to the travel-time algorithms used in geophysics and radar imaging, such as those used 
in e.g. Valle {\it et al} (1999) and Skolnik (2001). 

We have also considered the instantaneous power flow from a transient source. In the near-field zone
power may flow towards the source for some time. Although, the time-integrated
power flow stays positive, a positive instantaneous power flow followed by a 
negative flow during a significant interval of time will induce an oscillating linear momentum and an
oscillating instantaneous light pressure. 

Finally, we have analyzed the local instantaneous energy velocity which is conveniently bounded 
between $-c$ and $c$. Its waveforms show that the energy velocities of the wavefront
and the end of a wave-packet are both exactly $c$, and that the subluminal and negative
values during extended time intervals are the feature of the near-field zone only. 
At the same time the energy velocity reveals that the power flow can become negative
for a short time even in the far-field zone. This happens when both the power flow 
and the energy are close to zero, and it remains to be seen whether this effect can be 
detected experimentally.       

On a theoretical level all these unusual effects and their experimental confirmation 
show the correctness of the time-domain radiation 
formula (\ref{eq:XTFormula})--(\ref{eq:XTFormulaH}). We have demonstrated how this formula follows from 
the Lorenz gauge-fixing between the classical scalar and vector
potentials and eventually reduces to the continuity equation between the
current and charge densities. 
These otherwise somewhat {\it ad hoc} procedures thus acquire additional
physical significance.    


\end{document}